\documentclass[twocolumn]{jetpl}

\usepackage{graphicx}
\usepackage{cite}
\usepackage{amsmath}
\usepackage{amsfonts}
\usepackage{color}

\DeclareMathOperator{\tr}{tr}

\begin{document}

\title{Cross-over behavior of disordered interacting two-dimensional electron
systems in a parallel magnetic field}
\rtitle{Cross-over behavior of disordered  two-dimensional
electron systems in a parallel magnetic field}
\sodtitle{Cross-over behavior of disordered  two-dimensional
electron systems in a parallel magnetic field}

\author{I.\,S.\,Burmistrov$^{1,2}$ and N.\,M.\,Chtchelkatchev$^{1,2}$
\thanks{e-mail: burmi@itp.ac.ru, nms@itp.ac.ru}}
\rauthor{I.\,S.\,Burmistrov and N.\,M.\,Chtchelkatchev}
\sodauthor{I.\,S.\,Burmistrov and N.\,M.\,Chtchelkatchev}

\dates{\today}{*}

\address{
$^{1}$ L.D. Landau Institute for Theoretical Physics, Kosygina
street 2, 117940 Moscow, Russia\\
$^{2}$ Department of Theoretical Physics, Moscow Institute of
Physics and Technology, 141700 Moscow, Russia
}

\abstract{We present analysis of the cross-over behavior of
disordered interacting two-dimensional electron systems in the
parallel magnetic field. Using the so-called cross-over one-loop
renormalization group equations for the resistance and
electron-electron interaction amplitudes we qualitatively explain
experimentally observed transformation of the temperature
dependence of the resistance from a reentrant (nonmonotonic)
behavior in relatively weak fields into an insulating-type
behavior in stronger fields. }


\maketitle

Disordered two-dimensional electron systems (2DES) have been in
the focus of experimental and theoretical research for several
decades~\cite{AFS}. The interest to 2DES has been renewed because
of the discovery of the metal-insulator transition (MIT) in a high
mobility silicon metal-oxide-semiconductor field-effect transistor
(Si-MOSFET)~\cite{Pudalov1}. During last decade an interesting
behavior of resistance and spin susceptibility has been found
experimentally not only in Si-MOSFET but in other 2D electron
systems~\cite{Review}. Recently, a major step toward the
theoretical proof for the MIT existence in 2DES has been made in
Ref.~\cite{LargeN}.

\begin{figure}[b]
\vspace{-1cm} \centerline{\includegraphics[width=70mm]{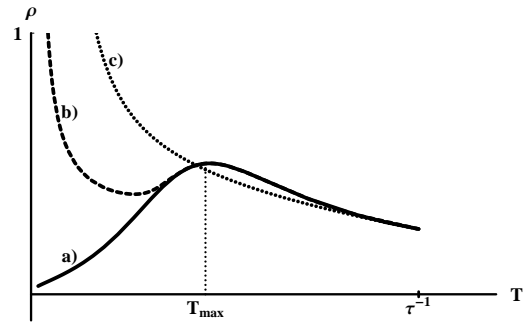}}
\caption{Fig.~\ref{Figure0}. The sketch of the temperature
dependence of the resistance in the presence of $B_\parallel$. The
curve a) corresponds to the case $B_\parallel=0$, the curve b) to
the case $g_L\mu_B B_\parallel \ll T_\textrm{max}$ and the curve
c) to the case $g_L\mu_B B_\parallel \gg T_\textrm{max}$
\label{Figure0}}
\end{figure}

If an electron density is higher than the critical one (metallic
phase) at low temperatures $T \ll \tau^{-1}$ [$\tau$ stands for
the transport mean-free path time] the increase of the resistance
with decreasing temperature is replaced by the drop as $T$ becomes
lower than some $T_\textrm{max}$ (see
Fig.~\ref{Figure0})~\cite{Review}. This nonmonotonic behavior of
the resistance has been predicted from the renormalization group
(RG) analysis of the interplay between disorder and
electron-electron interaction in 2DES~\cite{Finkelstein,FP}. As a
weak magnetic field $B_\parallel$ is applied parallel to 2DES the
decrease of the resistance is stopped at some temperature and the
resistance increases again~\cite{Pudalov2}. Further increase of
$B_\parallel$ leads to the monotonic growth of the resistance as
temperature is lowered.

In the Letter we present the theoretical explanation for this
striking behavior of the resistance in parallel magnetic field. We
demonstrate that it can be explained with the help of the RG
analysis of disorder and electron-electron interaction in 2DES in
the presence of the Zeeman splitting.

%
%
%
The presence of the parallel magnetic field introduces a new
energy scale $g_L\mu_B B_\parallel$ into the problem. Here $g_L$
and $\mu_B$ stand for the Land\'e factor and the Bohr magneton
respectively. The Zeeman splitting $g_L\mu_B B_\parallel$ sets the
cut-off for a pole in the diffusive modes (diffusons) with
opposite electron spin projections~\cite{Finkelstein}. In the
temperature range $g_L\mu_B B_\parallel\ll  T \ll \tau^{-1}$ this
cut-off is irrelevant and 2DEG behaves as if no parallel magnetic
field is applied. However, at lower temperatures $T \ll  g_L\mu_B
B_\parallel$ the diffusive modes
 with opposite electron spin projections do not
contribute and 2DES behaves as in the presence of the strong
parallel magnetic field $B_\parallel^{\infty}\sim
(g_L\mu_B\tau)^{-1}$. It is well known~\cite{Finkelstein} that in
this case the resistance monotonically grows as $T$ is lowered.
Therefore, if the parallel magnetic field is weak $g_L\mu_B
B_\parallel \ll T_\textrm{max}$, then the $T$-dependence of the
resistance manifests the reentrant behavior with maximum and
minimum. At higher parallel magnetic fields $g_L\mu_B B_\parallel
\gg T_\textrm{max}$ the insulating-type behavior of the resistance
restores such that it grows monotonically with decreasing $T$ (see
Fig.~\ref{Figure0}).

Below we shall illustrate this simple physical explanation of
non-monotonic behavior of the resistance by the so-called
cross-over RG equations (cf. Eqs.~\eqref{RGS1}-\eqref{RGS5}) for
the resistance and the electron-electron interaction amplitudes
that smoothly interpolate between the well-known cases of
$B_\parallel=0$ and strong field.

%
%
%
We consider two-dimensional interacting electrons with $n_v$
valleys in the presence of the quenched disorder and the parallel
magnetic field at low temperatures $T\ll \tau^{-1}$. We assume
that a magnetic field $B_\perp \gtrsim T/(D e)$ where $e$ and $D$
stand for the electron charge and diffusion coefficient
respectively is applied perpendicular to 2DES in order to suppress
the Cooper channel. We suppose that both the inverse intervalley
scattering time and the valley splitting are much less than the
temperature.

Following Finkelstein~\cite{Finkelstein}, the effective quantum
theory of disordered interacting two-dimensional electrons is
given in terms of the generalized non-linear $\sigma$-model
involving unitary matrix field variables
$Q^{\alpha_1\alpha_2;\zeta_1\zeta_2}_{mn}(\mathbf{r})$ which obey
the constraint $Q^2(\mathbf{r})=1$. Here the integers
$\alpha_i=1,2,\dots,N_r$ denote the replica indices, $m, n$
correpond to the discrete set of the Matsubara frequencies
$\omega_n=\pi T (2n+1)$. The integers $\zeta_i=\pm 1,\pm
2,\dots,\pm n_v$ are the combined spin and valley indices. The
effective action is given as
\begin{gather}
S = \int d^2\mathbf{r} \left
(\mathcal{L}_\sigma+\mathcal{L}_F+\mathcal{L}_{B_\parallel}+\mathcal{L}_h\right
). \label{SDef}
\end{gather}
Here $\mathcal{L}_\sigma$ is the free electron
part~\cite{FreeElectrons}
\begin{gather}
\mathcal{L}_\sigma=\frac{\sigma}{16 n_v}\tr (\nabla Q)^2,
\label{SDefSigma}
\end{gather}
where $\sigma$ denotes the mean-field conductance in units $e^2/h$
with $h$ being the Plank constant and symbol $\tr$ is the trace
over Matsubara, replica, spin and valley indices. The
$\mathcal{L}_F$ involves the electron-electron interaction
amplitudes which describe the scattering on small ($\Gamma$) and
large ($\Gamma_2$) angles and the quantity $z$ (originally
introduced by Finkelstein~\cite{Finkelstein}) which is responsible
for the specific heat renormalization~\cite{CasDiCas},
\begin{gather}
\mathcal{L}_F = 4\pi T z \tr\eta(\Lambda-Q)-\pi T\Gamma
\sum_{\alpha n}\tr I_n^\alpha Q\tr I_{-n}^\alpha Q \notag
\\
+ \pi T\Gamma_2\sum_{\alpha n}(\tr I_n^\alpha Q)\otimes(\tr
I_{-n}^\alpha Q)+ 2\pi T z \tr\eta \Lambda.\label{SDefF}
\end{gather}
Here $\tr A \otimes \tr B = A^{\alpha\alpha;\zeta_1\zeta_2}_{nn}
B^{\beta\beta;\zeta_2\zeta_1}_{mm}$ and the matrices $\Lambda$,
$\eta$ and $I_k^\gamma$ are given as
\begin{gather}
\Lambda^{\alpha\beta;\zeta_1\zeta_2}_{nm} =
\mathrm{sign}\,(\omega_n)
\delta_{nm}\delta^{\alpha\beta}\delta^{\zeta_1\zeta_2}, \notag\\
\eta^{\alpha\beta;\zeta_1\zeta_2}_{nm} = n
\delta_{nm}\delta^{\alpha\beta}\delta^{\zeta_1\zeta_2},
\label{matrices_def}\\
(I_k^\gamma)^{\alpha\beta;\zeta_1\zeta_2}_{nm} =
\delta_{n-m,k}\delta^{\alpha\gamma}\delta^{\beta\gamma}\delta^{\zeta_1\zeta_2}.
\notag
\end{gather}
In the presence of the parallel magnetic field $B_\parallel$ the
Zeeman splitting should be taken into account~\cite{Finkelstein}
\begin{gather}
\mathcal{L}_{B_\parallel}=- i z_2 g_L\mu_B B_\parallel \tr \tau_z
Q + \frac{n_v g_L^2\mu_B^2z_2}{2\pi T} N_r B_\parallel^2.
\label{SDefB}
\end{gather}
Here $z_2=z+\Gamma_2$ and the Pauli matrix $\tau_z$ is defined as
\begin{equation}
(\tau_z)_{nm}^{\alpha_1\alpha_2;\zeta_1\zeta_2} =
\mathrm{sign}\,(\zeta_1)
\delta_{nm}\delta^{\alpha\beta}\delta^{\zeta_1\zeta_2} .
\label{DefTauz}
\end{equation}
We mention that the last term in $\mathcal{L}_{B_\parallel}$
corresponds to the Fermi-liquid spin susceptibility. Finally, the
term
\begin{gather}
\mathcal{L}_h =-\frac{\sigma h^2}{4n_v}\tr\Lambda Q\label{SDefh}
\end{gather}
is not a part of the theory but we shall use it later on as a
convenient infrared regulator of the theory.

The action~\eqref{SDef} involves the matrices which are formally
defined in the infinite Matsubara frequency space. In order to
operate with them we have to introduce a cut-off for the Matsubara
frequencies. Then the set of rules which is called
$\mathcal{F}$-algebra can be established~\cite{Unify}. At the end
of all calculations we tend the cut-off to infinity.

The theory~\eqref{SDef} should be supplemented by the important
constraint that the combination $z+\Gamma_2-2n_v\Gamma$ remains
constant in the course of the RG flow. Physically, it corresponds
to the conservation of the number of particles~\cite{Finkelstein}.
In the special case of the Coulomb interaction which is of the
main interest for us in the paper the relation
$z+\Gamma_2-2n_v\Gamma=0$ holds and the action~\eqref{SDef} with
$B_\parallel=0$ is invariant under a global rotation of the matrix
$Q$ ($\mathcal{F}$-invariance)~\cite{Unify}.

%
%
%
The most significant physical quantities in the theory comprising
the complete information on its low-energy dynamics are the
physical observables $\sigma^\prime$, $z_2^\prime$ and $z^\prime$
associated with the mean-field parameters $\sigma$, $z_2$ and $z$
of the action~\eqref{SDef}. The quantity $\sigma^\prime$ is the
conductance of 2DES as one can obtain from a linear response to
electromagnetic field, $n_v g_L^2\mu_B^2 z_2^\prime/\pi$ is the
spin susceptibility of 2DES, and $z^\prime$ is related with the
specific heat of 2DES~\cite{Unify}. Extremely important to remind
that the observable parameters $\sigma^\prime$, $z_2^\prime$ and
$z^\prime$ are precisely the same as those defined by the
background field procedure~\cite{BurmistrovPruisken}.

The conductance $\sigma^\prime$ is expressed in terms of the
current-current correlations as~\cite{Unify}
\begin{gather}
\sigma^\prime = -\frac{\sigma}{8n_v n}\left
\langle\tr[I_{n}^\alpha, Q][I_{-n}^\alpha, Q] \right
\rangle\hspace{2.7cm}\,{} \label{SigmaODef}
\\
+ \frac{\sigma^2}{16n_v^2 n d} \int
d^d\mathbf{r}^\prime\langle\langle \tr I_n^\alpha
Q(\mathbf{r})\nabla Q(\mathbf{r}) \tr I_{-n}^\alpha
Q(\mathbf{r}^\prime)\nabla Q(\mathbf{r}^\prime)\rangle \rangle
\notag
\end{gather}
where the limit $n\to 0$ is assumed and $d$ denotes the dimension.
Here and from now onwards the expectations are defined with
respect to the theory~\eqref{SDef}. The observable $z_2^\prime$ is
given by~\cite{Finkelstein}
\begin{gather}
z_2^\prime =\frac{\pi}{n_v
(g_L\mu_B)^2N_r}\frac{\partial^2\Omega}{\partial B_\parallel^2}
,\label{z2Def}
\end{gather}
where $\Omega$ denotes the thermodynamic potential of the unit
volume. A natural definition of $z^\prime$ is obtained through the
derivative of $\Omega$ with respect to $T$~\cite{Unify},
\begin{gather}
z^\prime = \frac{1}{2\pi \tr \eta \Lambda} \frac{\partial
}{\partial T}\frac{\Omega}{T}.\label{zDef}
\end{gather}

%
%
%
To define a theory for perturbative expansions we use the
``square-root'' parameterization
\begin{gather}
Q = W+\Lambda\sqrt{1-W^2},\qquad W =
  \begin{pmatrix}
    0 & w \\
    w^\dag & 0
  \end{pmatrix} .
\end{gather}
The action~\eqref{SDef} can be written as an infinite series in
the independent fields
$w_{n_1n_2}^{\alpha_1\alpha_2,\zeta_1\zeta_2}$ and
$w_{n_4n_3}^{\dag \alpha_1\alpha_2,\zeta_1\zeta_2}$. We use the
convention that Matsubara indices with odd subscripts $n_1, n_3,
\dots$ run over non-negative integers whereas those with even
subscripts $n_2, n_4, \dots$ run over negative integers.

The propagators can be written in the following form
\begin{gather}
\langle w_{n_1n_2}^{\alpha_1\alpha_2;\zeta_1,\zeta_2}(p) w_{n_4
n_3}^{\dag\alpha_4\alpha_3;\zeta_4\zeta_3}(-p) \rangle =
\frac{8n_v}{\sigma}
\delta^{\alpha_1\alpha_3}\delta^{\alpha_2\alpha_4} \label{Prop}
\\
\times \delta_{n_{12},n_{34}} \Bigl
[\delta_{n_1,n_3}\delta^{\zeta_1\zeta_3}\delta^{\zeta_2\zeta_4}
D_p(\omega_{12},i\omega_{12}^{B})\hspace{0.6cm}\,{}\notag
\\
-\frac{16 \pi n_v  T z\gamma_2}{\sigma}
\delta^{\alpha_1\alpha_2}\delta^{\zeta_1\zeta_3}\delta^{\zeta_2\zeta_4}
 D(\omega_{12},i \omega_{12}^{B})D^t_p(\omega_{12},i \omega_{12}^{B})\notag
 \\
  +
\frac{8 \pi T z(1-\alpha+\gamma_2)}{\sigma}
\delta^{\alpha_1\alpha_2}\delta^{\zeta_1\zeta_2}\delta^{\zeta_3\zeta_4}
D^s(\omega_{12})D^t_p(\omega_{12}) \Bigr ] ,\notag
\end{gather}
where we introduce the notations $z\alpha =
z+\Gamma_2-2n_v\Gamma$, $\gamma_2=1+\Gamma_2/z$,
$\omega_{12}=16n_v\pi Tz(n_1-n_2)/\sigma$,
$\omega_{12}^B=\omega_B(\mathrm{sign}\,\zeta_1-\mathrm{sign}\,\zeta_2)/2$
with $\omega_B=8n_v z_2 g_L\mu_B B_\parallel/\sigma$ and
\begin{gather}
D_p(\omega,x) = [p^2+h^2+\omega + x]^{-1},\notag
\\
D^s_p(\omega)= [p^2+h^2+\alpha \omega]^{-1},\notag
\\
D^t_p(\omega,x)=[p^2+h^2 +(1+\gamma_2)\omega+x]^{-1},\notag
\\
D_p(\omega) \equiv D_p(\omega,0),\qquad  D_p^t(\omega) \equiv
D_p^t(\omega,0).
\end{gather}
%

%
%
%
The standard one-loop analysis for the physical observables
$\sigma^\prime$, $z_2^\prime$ and $z^\prime$ performed at $T=0$
yields
\begin{gather}
\sigma^\prime = \sigma +\frac{8}{d}\int \frac{d^dp\,p^2}{(2\pi)^d}
\int_0^\infty d \omega \Bigl [ (2n_v^2-1)\gamma_2
D_p^2(\omega)D_p^t(\omega) \notag\\ +2n_v^2\gamma_2\Real
D_p^2(\omega,i\omega_B)D_p^t(\omega,i\omega_B)   -(1-\alpha)
D_p^2(\omega)D_p^s(\omega) \Bigr ] ,\notag
\end{gather}
\begin{gather}
z_2^\prime = z_2 +\frac{8n_v^2}{\sigma}z_2(1+\gamma_2) \Real\int
\frac{d^dp}{(2\pi)^d}\int_0^\infty d\omega
\hspace{1.4cm}\,{}\notag
\\ \times \Bigl[D_p^2(\omega,i\omega_B)-D_p^{t2}(\omega,i\omega_B)\Bigr ]
,\notag
\\
z^\prime = z  +\frac{2z}{\sigma}  \int \frac{d^dp}{(2\pi)^d}\Bigl
[ \alpha D_p^s(0)-2n_v^2 \Real D_p(0,i\omega_B)\hspace{1cm}\,{}
\notag\\- 2n_v^2 D_p(0)+(2n_v^2-1)(1+\gamma_2) D_p^t(0) \notag
\\
+2n_v^2(1+\gamma_2)\Real D_p^t(0,i\omega_B)  \Bigr ]
.\label{1loop}
\end{gather}
In what follows we shall employ the dimensional regularization
scheme with $d=2+2\epsilon$. Evaluating the momentum and frequency
integrals in Eqs.~\eqref{1loop} for the case of the Coulomb
interaction ($\alpha=0$) we obtain for the physical observables
$\rho^\prime = 4 \Gamma(1-\epsilon)(4\pi)^{-d/2}/\sigma^\prime$,
$z_2^\prime$ and $z^\prime$
\begin{gather}
\frac{1}{\rho^\prime} = \frac{1}{\rho}+ \frac{
h^{2\epsilon}}{\epsilon} \Bigl [1+g_t(\gamma_2) -2n_v^2
g_t(\gamma_2) (1+f_\epsilon(\omega_B/h^{2}))\Bigr ], \notag
\\
z_2^\prime=z_2\left [ 1-\frac{2n_v^2\gamma_2\rho\,
h^{2\epsilon}}{\epsilon}f_\epsilon(\omega_B/h^{2})\right ],
\hspace{1.5cm}\,{}\label{POeps}\\
z^\prime=z\Bigl [1+ \frac{\rho\, h^{2\epsilon}}{2\epsilon}\Bigl
(1-\gamma_2(2n_v^2-1)+2n_v^2 \gamma_2f_\epsilon(\omega_B/
h^{2})\Bigr )\Bigr ], \notag
\end{gather}
where $\rho=4\Gamma(1-\epsilon)(4\pi)^{-d/2}/\sigma$,
$f_y(x)=\Real(1+i x)^y$ and
\begin{gather}
g_t(\gamma_2)=\frac{1+\gamma_2}{\gamma_2}\ln(1+\gamma_2)-1,
\end{gather}
The standard minimal subtraction scheme is the
$\omega_B$-independent and, therefore, is unable to treat the
limits of vanishing ($\omega_B=0$) and strong
($\omega_B\to\infty$) magnetic fields simultaneously. In order to
avoid this problem and obtain the renormalization that correctly
describes the two limiting cases of $\omega_B=0$ and
$\omega_B\to\infty$ we shall use the so-called cross-over
renormalization scheme~\cite{Amit}.
%
%
\begin{table*}[t]
\caption{Table. The function $F(\gamma_2)$ for the cases of
vanishing and strong magnetic field. Symbol
$\text{li}_n(x)=\sum_{k=1}^\infty x^k/k^n$ denotes the
polylogarithmic function.}
\begin{tabular}{c|c}
$f_c$& $F(\gamma_2)$
\\
\hline&\\
$0$ & $\frac{8n_v^2\gamma_2}{1+\gamma_2}
+(4n_v^2-1)\ln^2(1+\gamma_2)+2(4n_v^2-1) \mathrm{li}_2(-\gamma_2)
$
\\
$1$ & $   \begin{cases}
   2g_t\left(\frac{-1}{2n_v^2}\right )\ln
[1-(2n_v^2-1)\gamma_2] +2(2n_v^2-1) \Bigl
[\mathrm{li}_2(-\gamma_2)-\mathrm{li}_2\left
(\frac{1-(2n_v^2-1)\gamma_2}{2n_v^2}\right) \Bigr ] +2 \ln
(1+\gamma_2), & \gamma_2<\gamma_2^\star\\
   2g_t\left(\frac{-1}{2n_v^2}\right )\ln \left
[1-\frac{1}{(2n_v^2-1)\gamma_2}\right ] +2(2n_v^2-1) \Bigl
[\mathrm{li}_2\left
(1-\frac{1}{(2n_v^2-1)\gamma_2}\right)-\mathrm{li}_2\left(\frac{(2n_v^2-1)\gamma_2-1}{2n_v^2\gamma_2}\right
) \Bigr ] +2 \ln \frac{1+\gamma_2}{\gamma_2},  & \gamma_2\geqslant
\gamma_2^\star
  \end{cases}
$\\
\end{tabular}
\end{table*}

%
%
%
The parameter $h$ naturally sets the momentum scale at which the
bare parameters $\rho$, $z_2$ and $z$ of the action~\eqref{SDef}
are defined. The physical observables $\rho^\prime$, $z_2^\prime$
and $z^\prime$ correspond to the momentum scale
$h^\prime$~\cite{Pruisken,BurmistrovPruisken2} which is determined
as
%
$\sigma^\prime h^\prime \tr 1 = \sigma h \tr \Lambda \langle
Q\rangle$.
%
By using the relation $h^\prime=h[1+O(\rho)]$ we can substitute
$h^\prime$ for $h$ in Eqs.~\eqref{POeps}. Since the bare
parameters $\rho$, $z_2$, and $z$ are independent of $h^\prime$,
we obtain from Eqs.~\eqref{POeps} the following one-loop
cross-over RG equations in two dimensions ($\epsilon=0$)
\begin{eqnarray}
\frac{d\rho^\prime}{d \eta} &=&  [a_0(\gamma_2^\prime)+ f_c
a_1(\gamma_2^\prime)]\rho^{\prime 2} ,\label{RGS1}
\\
\frac{d \gamma_2^\prime}{d \eta}&=&-\rho^\prime \Bigl [
b_0(\gamma_2^\prime)+f_c b_1(\gamma_2^\prime)\Bigr ] ,
\label{RGS2}
\\
\frac{d \ln z^\prime}{d \eta}&=&-\rho^\prime \Bigl
[c_0(\gamma_2^\prime)+f_c c_1(\gamma_2^\prime)\Bigr ],
\label{RGS3}\\ \frac{d f_c}{d \eta}&=&-4 f_c(1-f_c),\quad
f_c(0)=\mathcal{B}^2/(1+\mathcal{B}^2).\label{RGS4}
\end{eqnarray}
Here we have introduced $\gamma_2^\prime=z_2^\prime/z^\prime -1$
and variable $\eta = \ln l h^\prime$ where $l\propto h^{-1}$
corresponds to a mean-free path length at which, physically, the
bare parameters of the action~\eqref{SDef} are defined and
\begin{gather}
a_0(\gamma_2^\prime) = -2\left [ 1-
(4n_v^2-1)g_t(\gamma_2^\prime)\right ],\,b_0(\gamma_2^\prime) =
(1+\gamma_2^\prime)^2, \notag
\\
c_0(\gamma_2^\prime)=(4n_v^2-1)\gamma_2^\prime-1,\quad
a_1(\gamma_2^\prime) = -4n_v^2g_t(\gamma_2^\prime), \notag
\\
b_1(\gamma_2^\prime)
=
-2n_v^2\gamma_2^\prime(1+\gamma_2^\prime), \quad
c_1(\gamma_2^\prime)= -2n_v^2\gamma_2^\prime. \label{RGS5}
\end{gather}
The cross-over parameter $f_c$ is expressed via the quantity
$\mathcal{B}=\omega_B l^2$. Eqs.~\eqref{RGS1}-\eqref{RGS3}
smoothly interpolate between the known results for the cases
$f_c=0$ and $f_c=1$~\cite{Finkelstein,FP}. They were derived under
assumption that $\bar{\rho}^\prime \ll 1$. From here onwards we
omit the `prime' sign for convenience.

In general, the function $f_c(\eta)$, i.e., the right hand side of
Eq.~\eqref{RGS4}, is not universal. It depends on a method
employed to derive the cross-over RG equations. The only universal
properties of Eq.~\eqref{RGS4} are the existence of two fixed
points $f_c=0$ and $f_c=1$. Fortunately, the qualitative features
of the RG flow for Eqs.~\eqref{RGS1}-\eqref{RGS5} are independent
of the choice of $f_c(\eta)$ provided it smoothly interpolates
between $f_c=0$ at $\mathcal{B} e^{-2\eta}\ll 1$ and $f_c=1$ at
$\mathcal{B} e^{-2\eta}\gg 1$.

\begin{figure}[b]
\vspace{-0.5cm} \centerline{\includegraphics[width=70mm]{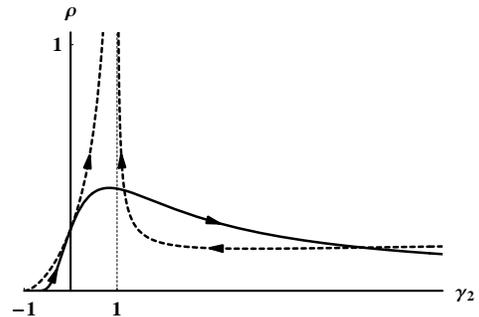}}
\caption{Fig.~\ref{Figure1}. RG flow diagram in $\rho$ versus
$\gamma_2$ for $f_c=0$ (solid line) and for $f_c=1$ (dashed lines)
with $n_v=1$. The arrows indicate the direction towards the
infrared $L\to\infty$ ($\eta\to -\infty$). See
text\label{Figure1}}
\end{figure}

%
%
%
Let us start the analysis of Eqs.~\eqref{RGS1}-\eqref{RGS3} from
the case $f_c=0$. Then we find $\rho(\eta)
\propto\exp{F(\gamma_2(\eta))}$ where the function $F(\gamma_2)$
is given in Table. For all values of $n_v$ $\gamma_2$ increases
monotonically  with decreasing $\eta$ and diverges at $\eta_c
=\Upsilon(\infty;\bar{\rho},\bar{\gamma}_2)$ where
\begin{gather}
\Upsilon(\gamma_2;\bar{\rho},\bar{\gamma}_2)=-
\frac{1}{\bar{\rho}} e^{F(\bar{\gamma}_2)}
\int_{\bar{\gamma}_2}^{\gamma_2}
\frac{du}{b(u)}e^{-F(u)}.\label{etac0}
\end{gather}
with $b(u)\equiv b_0(u)$, $\bar{\rho}=\rho(0)$ and
$\bar{\gamma}_2=\gamma_2(0)$. We present the RG flow diagram for
$n_v=1$ in Fig.~\ref{Figure1}. The RG flow is qualitatively the
same for all values of $n_v$. As $\gamma_2$ monotonous increases
an initial growth of $\rho$ changes into a decline, i.e., 2DES
remains in the metallic phase ($\rho\to 0$) at large lengthscales
($\eta\to -\infty$).

%

In the opposite case $f_c=1$ the integration of
Eqs.~\eqref{RGS1}-\eqref{RGS3} yields $\rho(\eta)\propto
\exp{F(\gamma_2(\eta))}$ with the function $F(\gamma_2)$ presented
in Table. If $\bar{\gamma}_2<\gamma_2^\star=(2n_v^2-1)^{-1}$
($\bar{\gamma}_2>\gamma_2^\star$) $\gamma_2$ increases (decreases)
monotonous as $\eta$ diminishes. It reaches finally the value
$\gamma_2^\star$ at
$\eta_c=\Upsilon(\gamma_2^\star;\bar{\rho},\bar{\gamma}_2)$ where
the function $\Upsilon(\gamma_2;\bar{\rho},\bar{\gamma}_2)$ is
given by Eq.~\eqref{etac0} with $b(u)\equiv b_0(u)+b_1(u)$. We
mention that simultaneously $\rho(\eta)$ diverges at $\eta_c$. We
plot the RG flow diagram with $n_v=1$ in Fig.~\ref{Figure1}.
Unlike the case $f_c=0$, 2DES is in the insulating phase
($\rho\to\infty$) at large lengthscales for $f_c=1$, $\gamma_2$
being quenched to $\gamma_2^\star$.

In the intermediate case $0< f_c < 1$ the cross-over RG
Eqs.~\eqref{RGS1}-\eqref{RGS5} can be solved only numerically. If
the parallel magnetic field is smaller than some $\mathcal{B}_{X}$
which is a function of $\bar{\rho}$ and $\bar{\gamma}_2$ the
resistance has the reentrant behavior with the maximum and the
minimum at some values of $\eta$ (see Fig.~\ref{Figure0}). At
$\mathcal{B}>\mathcal{B}_{X}$ the insulator-type behavior of the
$\rho(\eta)$ that one expects for $\mathcal{B}\to\infty$ is
restored. For $\bar{\gamma}_2<\gamma_2^\star$ the function
$\gamma_2(\eta)$ has the maximum whereas
$z_2(\eta)=z(\eta)(1+\gamma_2(\eta)$ (spin susceptibility) has the
minimum. Their behavior does not change qualitatively when the
parallel magnetic field passes through the value $\mathcal{B}_X$.
For $\bar{\gamma}_2>\gamma_2^\star$ the maximum in the function
$\gamma_2(\eta)$ disappears for $\mathcal{B}>\mathcal{B}_{X}$
whereas $z_2(\eta)$ increases monotonous for all values of
$\mathcal{B}$. As expected, the cross-over field $\mathcal{B}_X$
can be estimated as $\mathcal{B}_X \sim \exp(2\eta_\textrm{max})$
where $\eta_\textrm{max} =
\Upsilon(\gamma_2^\textrm{max};\bar{\rho},\bar{\gamma}_2)$
determines the position of the maximum on the curve $\rho(\eta)$
at $f_c=0$. Here $\gamma_2^\textrm{max}$ is given as the root of
equation $a_0(\gamma_2)=0$.

%
%
%
We have not considered above the contribution to the one-loop RG
equations from the particle-particle (Cooper) channel. It can be
shown~\cite{AA} that the Zeeman splitting due to applied parallel
magnetic field does not affect it in the one-loop approximation.
Therefore, the Cooper-channel contribution to the RG equations can
be taken into account by the substitution of $a_0(\gamma_2)-2 n_v$
for $a_0(\gamma_2)$~\cite{FP}.  As one can check, the behavior of
$\rho(\eta)$, $\gamma_2(\eta)$ and $z_2(\eta)$ in this case
remains qualitatively the same as discussed above.

At zero temperature, $h^\prime$ plays the role of the inverse
lengthscale $L$ which is physically nothing else than a sample
size of 2DES. If $L \gg \sqrt{D/T}$ the temperature behavior of
the physical observables can be found from the cross-over RG
Eqs.~\eqref{RGS1}-\eqref{RGS3} stopped at the inelastic length
$L_\textrm{in}$. Formally, it means that one should substitute
$\eta_T= \frac{1}{2}\ln T l^2/D$ for $\eta$ with the help of the
following equation$\frac{d \eta_T}{d \eta} =1-\frac{d\ln z}{2
d\eta}$ ~\cite{Euro}. The temperature dependence of the resistance
in the presence of $B_\parallel$ obtained from $\rho(\eta)$ in
this way agrees qualitatively with the experimental results of
Ref.~\cite{Pudalov2}. The detailed comparison between the theory
and the experimental data on the resistance in the parallel
magnetic field will be presented in the following
Letter~\cite{FolLetter}.

Also, we remind that the physical observables that we use
throughout the paper are \emph{ensemble averaged}. Even in a
macroscopic sample of size $L \gg L_\textrm{in}$ they are
different from the \emph{measured} resistance $\rho(T)$ and spin
susceptibility $z_2(T)$ due to statistically independent
fluctuations of local conductance and electron-electron amplitudes
in blocks of the size $L_\textrm{in}$~\cite{Cohen}.

Finally, we mention that the symmetry-breaking strain applied to
2DES with two-valleys~\cite{strain} should affect the temperature
dependence of the resistance in similar way as $B_\parallel$.

%
%
%
In summary, we have presented the analysis of the cross-over
behavior of disordered interacting 2DES in the parallel magnetic
field. The one-loop cross-over RG equations that smoothly
interpolate between the two well-known limiting cases of vanishing
and strong parallel magnetic field allow us to explain
qualitatively the reentrant (nonmonotonic) behavior of the
resistance of 2DES as a function of $T$ in the presence of a
relatively weak parallel magnetic field as well as its insulating
behavior at stronger fields.

%
%
%
We would like to thank D.A.\,Knyazev, O.E.\,Omelyanovsky, and
V.M.\,Pudalov for the detailed discussions of their experimental
data that motivated us for the present research. One of the
authors (ISB) is grateful to A.M.M.\,Pruisken for discussions on
the cross-over RG. The financial support from the Russian Ministry
of Education and Science, Council for Grants of the President of
Russian Federation, RFBR, Program of RAS "Quantum Macrophysics",
Russian Science Support Foundation and CRDF is acknowledged.

\end{document}